# Climbing Mont Blanc —
## A Training Site for Energy Efficient Programming on Heterogeneous Multicore Processors

Lasse Natvig, Torbjørn Follan, Simen Støa, and Sindre Magnussen

Department of Computer and Information Science (IDI)
Norwegian University of Science and Technology (NTNU)
Lasse@computer.org

Antonio Garcia Guirado
ARM Norway

## ABSTRACT
Climbing Mont Blanc (CMB) is an open online judge used for training in energy efficient programming of state-of-the-art heterogeneous multicores. It uses an Odroid-XU3 board from Hardkernel [15] with an Exynos Octa processor and integrated power sensors. This processor is three-way heterogeneous containing 14 different cores of three different types. The board currently accepts C and C++ programs, with support for OpenCL v1.1, OpenMP 4.0 and Pthreads. Programs submitted using the graphical user interface are evaluated with respect to time, energy used, and energy-efficiency (EDP). A small and varied set of problems are available, and the system is currently in use in a medium sized course on parallel computing at NTNU. Other online programming judges exist, but we are not aware of any similar system that also reports energy-efficiency.

## Categories and Subject Descriptors
B.8.2 [**Performance and reliability**]: Performance Analysis and Design Aids, D.1.3 [**Programming techniques**]: Concurrent Programming – *Parallel programming,* K.3.2 [**Computers and Education**]: Computer and Information Science Education – *Parallel Computer science education programming.*

## General Terms
Algorithms, Measurement, Performance, Experimentation.

## Keywords
Heterogeneous multicore, Exynos processor, energy efficient programming, programming competition.

## 1. INTRODUCTION AND MOTIVATION
*Climbing Mont Blanc* (CMB) is an open online judge facilitating training and competitions in energy efficient programming of state-of-the-art heterogeneous multicores. The project name is inspired by the European research project Mont Blanc that started in October 2011 [13]. The Mont Blanc project aims to design a new type of computer architecture for future supercomputers using energy efficient processors from the embedded systems market and have built several prototypes using Samsung Exynos processors [5]. Exynos is a family of heterogeneous multicores or multi processor system on chip (MPSoC). In the CMB system, we use a similar processor and we envision competing programmers are *climbing* on the ranking list (Mont Blanc) of the submitted programs evaluated by CMB.

The current version of CMB uses an *Odroid-XU3* board from Hardkernel [15] with an Exynos 5 Octa (Exynos 5422) processor. It consists of four ARM Cortex-A15 and four ARM Cortex-A7 cores, and an ARM Mali-T628 GPU with six cores. The board has integrated power sensors that we use to measure the energy use of executed programs. CMB currently supports programs written in C++ and C, with support for OpenCL 1.1, OpenMP 4.0 and Pthreads. The Computer Architecture and Design group at IDI, NTNU (CARD), hosts the system and it is available for free trial use at https://climb.idi.ntnu.no. Additional information about the project will appear at https://www.ntnu.edu/idi/card/cmb.

Although the A7 and A15 CPUs run the same ISA they have very different performance characteristics, and with the six cores in the Mali GPU the chip can informally be called a *3-way heterogeneous multicore with 14 cores*. The four A7 and four A15 CPUs follows the ARM big.LITTLE heterogeneous computing architecture developed by ARM Holdings, coupling the slower, low-power A7 processor cores with the more powerful and power-hungry A15 cores. The Exynos 5422 provides *global task scheduling* that enable the use of all physical cores at the same time. Thus, programmers must decide not only the degree of parallelism but also how to schedule tasks to an optimal combination of the A7s, the A15s, or the GPUs running OpenCL code. *These additional possibilities add significantly to the complexity of programming, but makes it possible to achieve improved energy efficiency* and the right balance between low energy use and the performance required by the application user in a given scenario.

There is a large need for programmers being skilled in developing software for these heterogeneous multicores. The Exynos 5422 was used in the popular mobile phone Samsung Galaxy S5 [5], and the newer Exynos 7 Octa 7420 with similar heterogeneity is used in the new high-volume Smartphones Galaxy S6 and S6 Edge. The CMB system is a training site for this kind of programming, and we see energy efficient computing on one or more heterogeneous multicores of this kind as a key enabling technology for improved and new applications on handheld devices.

The effort to develop HPC systems from mobile SoCs in the Mont-Blanc project has resulted in several prototypes with growing energy-efficiency. *Tibidabo* was the first prototype (2011) [19]. It was a proof of concept of a cluster of mobile processors, deploying a full HPC software stack and uses Nvidia Tegra2 SoCs containing two ARM Cortex-A9 at 1GHz. *Pedraforca* (2013) was their first largescale prototype with a total of 78 Tegra3 compute nodes, each with 4 ARM Cortex-A9 at 1.3 GHz [13]. In February 2015, the Mont-Blanc project deployed the Mont-Blanc prototype containing 1080 Exynos 5250 SoCs organized in 72 compute blades over two racks [8, 13]. Exascale supercomputers containing huge numbers of energy-efficient heterogeneous multicores are now closer to reality thanks to these prototypes [18].

Programming such large heterogeneous computers, and especially exploiting their potential energy-efficiency effectively, poses numerous challenges. Climbing Mont Blanc aims to stimulate and help programmers focusing on those very challenges and therefore contribute to advances in the heterogeneous parallel programming field. Hence, CMB is an *educational project* for programming of mainstream processors used *in both embedded systems and HPC.*

The CMB project is further motivated by two other observations. Firstly, the last decade has shown an increase in the activity on online programming judges such as UVa Online Judge [20], PKU Judge Online [17], and KATTIS [10]. These show a huge activity — the numbers of uploaded programs per day are between three and four thousands for the first two. Secondly, it is a sad and well know fact that we currently have a rather high unemployment rate within the European Union. In June 2015, it was reported to be close to 10% [4]. It is our hope that the CMB system might stimulate unemployed programmers to train in the skills needed for future energy efficient computing of mainstream processors.

The rest of the paper is organized as follows. Section 2 briefly describes the system overview and some of the technical details of the current prototype. The main functionality and user modes are sketched in Section 3. The current prototype is fully functional and we have started to collect some user experience as described in Section 4, before we end the paper with discussion and future work.

## 2. SYSTEM OVERVIEW
### 2.1 System Architecture

The overall architecture of the current version of the Climbing Mont Blanc system is shown in Figure 1. It was developed by Torbjørn Follan and Simen Støa at NTNU, first as a specialization project in the autumn semester of 2014, and then as their Master Thesis "*Climbing Mont Blanc, A Prototype System for Online Energy Efficiency Based Programming Competitions on ARM* Platforms" [7]. CMB consist of three main parts, the *frontend*, the *server* and the *backend*.

- The *frontend* is the user interface, and handles all user related interactions like logging in and out of the system, browsing through the current list of available programming problems, submitting solutions for evaluation by CMB, or creating or joining a group. See Section 3 for a short description of the user functionality. The frontend is written using JavaScript (ECMAScript, 5$^{th}$ Edition) and AngularJS that is an open-source web application framework maintained by Google and individual developers [1]. AngularJS uses the Model-View-Controller pattern [11] and extends HTML with the possibility of dynamic views with two-way databinding. This makes the views update dynamically without refreshing the page, which leads to a smoother user experience. The code may also become more structured, as one can define own reusable components. The views are implemented with HTML5 [14] and CSS3 [3].

- The **server** is implemented as a REST API using Python Flask [6] and uses Gunicorn to handle multiple requests. The frontend sends requests to the server, and the server processes the requests and responds with JSON messages. All useful information, such as user and submission information, is stored in an SQLite database. SQLAlchemy is used as an Object Relational Mapper (ORM) on top of the database. The server is also responsible of submission, compiling and running of the submitted solutions on the backend (the XU3 board) by constantly checking the submission queue and communicate with the backend. The communication with the backend is via SSH. The server also runs a background cron job that checks the system state every 15 minutes, and an automatic mail is sent to the CMB team in case of system failure.

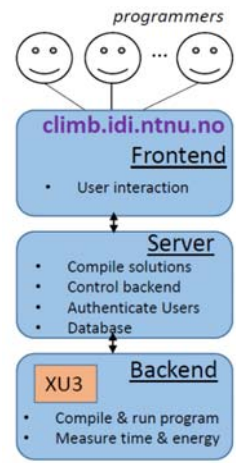

**Figure 1. Climbing Mont Blanc system architecture.**

- The *backend* is the Hardkernel Odroid-XU3 board. The current board runs Ubuntu Linux with some additional software to make it possible to monitor energy use. The server uses SSH to access the board as a user with minimal permissions. The backend compiles the programs uploaded by the server and runs them while measuring energy use. The observant reader might ask why we compile both at the server and the backend. The reasons are both the lack of a good enough cross-compiler and that we found it useful to do a quick compilation at the server to provide quick feedback to the users if there were any compilation errors in the uploaded program. The backend also does some processing on the measured energy statistics before it reports the output back to the server. Currently, only C and C++ programs can be compiled and ran on the backend. We compile with the g++ version 4.9.2, which has support for OpenCL v1.1, OpenMP 4.0 and Pthreads NTPL 2.19. Several plans for improvements of the backend side of CMB are sketched in Section 5.

The Odroid-XU3 is shown in Figure 2. Its main component, the Exynos 5422 was introduced above. As we can see from the figure, it is a typical evaluation board with numerous external interfaces and contacts. Of these, we only use the Ethernet port for access and the eMMC module for storage of the OS and the file system. Of most importance for us is the on board energy monitoring sensors making it possible to read out detailed energy consumption data by on-chip software.

### 2.2 Measuring Performance and Energy

Performance is measured as the wall clock time it takes to execute a submitted program. This is done by taking timestamps before and after the program executes. The energy is measured using the Hardkernel EnergyMonitor program, which is compatible with Odroid-XU3. This is started right before the uploaded program is executed and terminates when it is finished running. All energy readings are written to a file and processed before the result is reported. Both the energy and the Energy-Delay-Product (EDP) is calculated. The EDP is the main energy efficiency metric in the CMB system as it takes both running time and energy use into account. Before energy measuring starts, we clear the Linux OS cache in memory and stress the system to warm it up to a fixed starting temperature to ensure fair running time values and energy read outs [2] respectively among different submissions. Further details can be found in the Master Thesis of Follan and Støa [7].

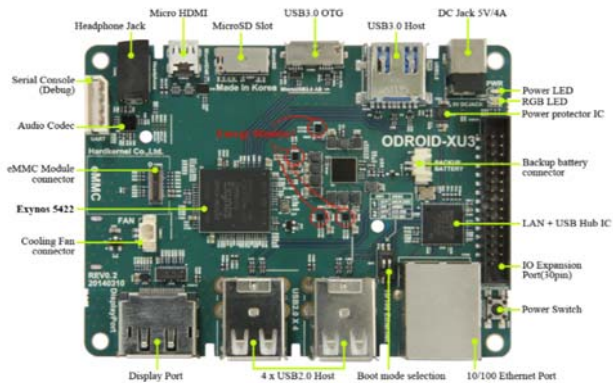

**Figure 2. The Odroid XU3 board (From [15])**

## 3. MAIN FUNCTIONALITY

The user interface of CMB is quite intuitive and due to space limitations, we will only introduce the main functionality in this paper. Figure 3 shows the CMB *home page*. To the left it shows the problems that are available to all users and by pressing the green button for one of the problems in the list, the problem page and its associated ranking lists are displayed. The *problem page* gives a precise specification of the problem and the format of the expected input and output. Further, it contains a *browse-button* for browsing your files and selecting a file for uploading it as a solution to the problem. CMB also displays a list of all your submissions for that problem and their results. The user can press a *Make-Private* or *Make-Public* button to control the visibility of these results.

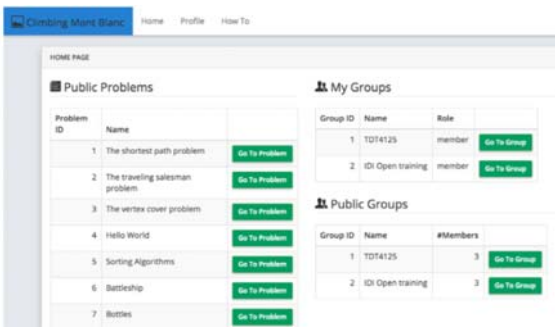

**Figure 3. Climbing Mont Blanc Home Page.**

As seen to the right in Figure 3, there is a list of public groups that the user can join, and it is displayed what groups the user already has joined. A *group* can be a specific course or a competition, and any user can define a new group and select which problems should be listed as included in that group.

Figure 4 shows two *highscore lists* for the shortest path problem. These lists are at the bottom of the corresponding problem page, and the *Choose Highscore button* allows the user to display the public high score or the list for a selected group where the problem is used. Also, by clicking on the heading Time, Energy or EDP — the list is sorted on the chosen criteria. (Some problems have an additional column displaying a "goodness"-metric as explained in Section 4.2).

**Figure 4. Public and Group Specific Highscore Lists.**

## 4. EARLY RESULTS AND EXPERIENCE

The first Climbing Mont-Blanc prototype was used in some smaller experiments during spring 2015, mainly in the course TDT4125 Algorithm Construction at NTNU. The system was slightly improved during summer 2015 and is now used for several compulsory exercises in the course TDT4200 parallel computing with more than 100 students.

### 4.1 Current Set of problems

The current set of problems are from three different sources.

- About 20 problems were ported from a set of training exercises used for the yearly IDI Open programming competition at NTNU. A few of these are easy and the rest span different difficulties from easy up to real algorithmic challenges even for a sequential solution.
- For the Algorithm Construction course, we defined four problems: the shortest path problem, the traveling salesman problem (TSP), the vertex cover problem and sorting.
- In the ongoing course in parallel computing there are three problems, one sequential focusing mainly on caching and two where the students are asked to program also the GPU using OpenCL.

### 4.2 Dealing with floating point calculations and approximation problems

When using problems involving floating point calculations in programming competitions it is a well-known fact [16] that special care must be taken since an exact checking byte for byte with "diff" or a similar command will not work. Therefore, CMB allows the problem designer to give a checker module that is a C++ program checking the solution of a submitted program. This allows the problem designer to specify the required level of numerical precision.

The use of NP-complete problems such as TSP advocates for allowing approximate solutions, and we have given the problem designer the possibility of a problem-specific goodness metric that is shown in the ranking list. Examples of its use are "total distance" for TSP and "cover size" for the vertex cover problem.

## 5. DISCUSSION AND FUTURE WORK

The very first user experience from the CMB system used for compulsory exercises in a medium sized course is gathered these days. We are not aware of any other online programming judges that measures energy-efficiency. The closest activity we have found is a Green Coding Contest given by Intel in 2014 [9].

In the shorter term, we will aim at getting more feedback from users and experience from operating the system. We will focus on maintaining high availability and observe the capacity to handle many submissions during peak hours before deadlines. This will continue into the spring semester of 2016 when we are going to use CMB in two courses at NTNU.

The system is open for other users, but currently without any user support. We will work on improving the documentation for users and expand the base of problems. A main activity the next 8 months will be development of improved functionality as well as scalability to make the system able to digest an expected increase in traffic. Figure 5 gives a sketch of current ideas for how to increase the scalability and to make future extensions on the backend side possible. We will allow multiple backends, and a *broker* will distribute the load on these to be able to serve more submissions per hour. Our first step will be to add more XU3 boards. The broker will cooperate with the server that will queue the submissions, and distribute submissions depending on the state of the XU3 boards.

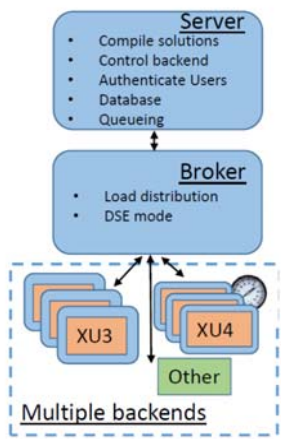

**Figure 5. Planned extensions of the CMB system.**

As also shown in Figure 5, we might add the newer Odroid-XU4 board [15] and other execution platforms. However, unfortunately the energy sensors being crucial for the CMB system is not any longer included. Since Odroid-XU3 is not any longer for sale [15], using the XU4 will require the integration of some external power measuring equipment as indicated in the figure. A tempting, and much easier alternative is to use a standard Intel desktop computer and its RAPL interface for energy read-outs, see e.g. our recent paper on this topic [12]. This would be to move away from the initial motivation of the project, though.

Offering different types of backends will also require redesign of the database and in the user interface (fronted). To add more languages such as Java and Python is also considered, and will require similar changes. We are also inspired by the recent improvements in functionality and GUI in the KATTIS system from KTH [10]. A more advanced and technically challenging extension is denoted DSE-mode (Design Space Exploration) in Figure 5. This is from early feedback from some users that want to interact with the system via a command-line interface or some other way to more easily loop through many different parameter settings for a single program being explored.

A more long-term goal for the CMB-project is to learn from submissions and try to distill knowledge about what is good practice for energy efficient programming on this kind of heterogeneous processors. In this context, we invite our colleagues to propose relevant problems with or without solution code.

## 6. ACKNOWLEDGMENTS
The Climbing Mont Blanc project has been supported by the Dept. of Computer and Information Science and by the ARM – CARD collaboration project, both at NTNU in Trondheim. The authors want to thank colleagues and students for their valuable contributions. In alphabetical order, they are Trond Aalberg, Jørn Amundsen, Christian Chavez, Arne Dag Fidjestøl, Rune Holmgren, Torje Hoås, Rune Erlend Jensen and Aleksander Rognhaugen.